\documentclass[pdflatex,sn-mathphys-num]{sn-jnl}

\usepackage{subcaption}
\usepackage{graphicx}%
\usepackage{tikz}
\usepackage[acronym]{acro}

\usepackage[section]{placeins} 

\usepackage{float}
\usepackage{multirow}%
\usepackage{amsmath,amssymb,amsfonts}%
\DeclareMathOperator{\diag}{diag}
\usepackage{amsthm}%
\usepackage[scr=rsfs]{mathalpha}%
\usepackage[title]{appendix}%
\usepackage{acro}%
\usepackage{braket}
\usepackage{xcolor}%
\usepackage{csquotes}
\usepackage{textcomp}%
\usepackage{physics}
\usepackage{manyfoot}%
\usepackage{booktabs}%
\usepackage{algorithm}%
\usepackage{algorithmicx}%
\usepackage{algpseudocode}%
\usepackage{listings}%
\usepackage[most]{tcolorbox}
\usepackage{float}
\usepackage{siunitx} 

\newcommand{\umu}{\ensuremath{\mathrm{\mu}}} 

\theoremstyle{thmstyleone}%

\usepackage[normalem]{ulem}

\theoremstyle{thmstyletwo}%

\theoremstyle{thmstylethree}%
\newcommand{\B}[1]{\mathbf{#1}}

\newcommand{\CommonAdvCaption}{%
Adversarial robustness with $\delta=N$ for (a) FGSM, (b) PGD, (c) DeepFool,
evaluated on classes $N\in\{4,6,8,10\}$ (atoms $=N$).
Top row (horizontal bars): mean $\Delta$ Accuracy per class,
$\overline{\Delta\mathrm{Acc}}_{N}
=\frac{1}{|\mathcal{E}|}\sum_{\varepsilon\in\mathcal{E}}
\big(\mathrm{Acc}^{\text{\ac{QRC}}+\text{\ac{MLP}}}_{N}(\varepsilon)
-\mathrm{Acc}^{\text{\ac{MLP}}}_{N}(\varepsilon)\big)$,
with $\mathcal{E}\subset[0.0,0.1]$.
Bottom row (line plots): Accuracy vs.\ $\varepsilon$ (solid: \ac{QRC}+\ac{MLP}; dashed: \ac{MLP}).
Positive bars indicate enhancement from \ac{QRC} (larger $\Delta$ Accuracy), while negative bars indicate degradation.%
}

\tcbuselibrary{skins} 

\newtcolorbox{redbox}{
  colback=yellow!10,    
  colframe=red!75,      
  boxrule=0.5pt,        
  arc=4pt,              
  left=4pt,             
  right=4pt,
  top=4pt,
  bottom=4pt,
  breakable             
}

\newtcolorbox{bluebox}{
  colback=yellow!10,    
  colframe=blue!75,      
  boxrule=0.5pt,        
  arc=4pt,              
  left=4pt,             
  right=4pt,
  top=4pt,
  bottom=4pt,
  breakable             
}

\DeclareAcronym{CD}{
short = CD,
long = constant detuning
}

\DeclareAcronym{FGSM}{
short = FGSM,
long = Fast Gradient Sign Method
}

\DeclareAcronym{GPU}{
short = GPU,
long = graphics processing unit
}

\DeclareAcronym{HPC}{
short = HPC,
long = high-performance computing
}

\DeclareAcronym{LSTM}{
short = LSTM,
long = long short-term memory
}

\DeclareAcronym{MLP}{
short = MLP,
long = multilayer perceptron
}

\DeclareAcronym{NISQ}{
short = NISQ,
long = noisy intermediate-scale quantum
}

\DeclareAcronym{PCA}{
short = PCA,
long = principal component analysis
}

\DeclareAcronym{PGD}{
short = PGD,
long = Projected Gradient Descent
}
\DeclareAcronym{RNN}{
short = RNN,
long = recurrent neural networks
}

\DeclareAcronym{QRC}{
short = QRC,
long = quantum reservoir computing
}

\DeclareAcronym{QPU}{
short = QPU,
long = quantum processing unit
}

\DeclareAcronym{QML}{
short = QML,
long = quantum machine learning
}

\DeclareAcronym{VQC}{
short = VQC,
long = variational quantum circuit
}

\begin{document}

\title{Towards Quantum Enhanced Adversarial Robustness with Rydberg Reservoir Learning}

\author*[1]{\fnm{Shehbaz} \sur{Tariq}}\email{shehbaz.tariq@uni.lu}
\equalcont{These authors contributed equally to this work.}
\author[2]{\fnm{Muhammad} \sur{Talha}}\email{talhazakir@khu.ac.kr}
\equalcont{These authors contributed equally to this work.}
\author[1]{\fnm{Symeon} \sur{Chatzinotas}}\email{symeon.chatzinotas@uni.lu}
\author[2]{\fnm{Hyundong} \sur{Shin}}\email{hshin@khu.ac.kr}

\affil*[1]{\orgdiv{Interdisciplinary Centre for
Security, Reliability, and Trust (SnT)}, \orgname{University of Luxembourg}, \orgaddress{ \postcode{L-1855}, \city{Luxembourg City}, \country{Luxembourg}}}

\affil[2]{\orgdiv{Department of Electronics and Information Convergence Engineering}, \orgname{Kyung  Hee University}, 
\orgaddress{\street{1732, Deogyeong-daero, Giheung-gu,}, \city{Yongin-Si}, \postcode{17104},  \country{Korea}}}

\abstract{\Ac{QRC} leverages the high-dimensional, nonlinear dynamics inherent in quantum many-body systems for extracting spatiotemporal patterns in sequential and time-series data with minimal training overhead. Although QRC inherits the expressive capabilities associated with quantum encodings, recent studies indicate that quantum classifiers based on variational circuits remain susceptible to adversarial perturbations. In this perspective, we investigate the first systematic evaluation of adversarial robustness in a QRC-based learning model. Our reservoir comprises an array of strongly interacting Rydberg atoms governed by a fixed Hamiltonian, which naturally evolves under complex quantum dynamics, producing high-dimensional embeddings. A lightweight multilayer perceptron serves as the trainable readout layer. We utilize the balanced datasets, namely \textsc{MNIST}, \textsc{Fashion-MNIST}, and \textsc{Kuzushiji-MNIST} as a benchmark for rigorously evaluating the impact of augmenting the quantum reservoir with an \Ac{MLP} in white-box adversarial attacks to assess its robustness. We demonstrate that this approach yields significantly higher accuracy than purely classical models across all perturbation strengths tested. This hybrid approach reveals a new source of quantum advantage and provides practical guidance for the secure deployment of machine learning models on quantum-centric supercomputing with near-term hardware.
}


\maketitle

\section*{Introduction}\label{sec1}
Classical computing architectures are increasingly constrained by the saturation of Moore's Law, reaching fundamental limits in terms of transistor density, energy efficiency, and linear scalability \cite{TW:17:IEEE_M_CSE}. Quantum computing has emerged as a promising alternative, offering to overcome these constraints through quantum mechanical principles, including superposition and interference, as well as non-classical correlations such as entanglement \cite{BISBDJBMN:18:NPh}. Consequently, quantum algorithms can outperform their classical counterparts in problems such as quantum simulation and specific sampling tasks, with rigorous demonstrations of the quantum advantage achieved in programmable superconducting processors \cite{AAABBBBBCBB:19:Nat,KEAWRRNWZT:23:Nat} and photonic quantum computers \cite{ZWDCPWLQWDH:20:Sci}. However, quantum processors remain limited by qubit count, error rates, and the challenges associated with quantum error correction and error mitigation \cite{P:18:Quantum}. These challenges limit the scalability of \ac{QML} models on near-term devices, underscoring the need for architectures that strike a balance between expressiveness and hardware feasibility.

Within QML, variational approaches such as \acp{VQC} have been widely studied across classification, generative modeling, and kernel methods, but suffer from limitations including vanishing gradients, barren plateaus, and the significant overhead of classical optimization loops \cite{MBSBN:18:Nat_Comm,SGSAM:23:JPA,SK:22:PRXQ,CLGDBF:25:Nat_Comm} These limitations motivate exploration of alternative models beyond variational approaches, with \ac{QRC} offering a particularly promising route.

QRC is inspired by classical extreme learning machines such as echo state networks and recurrent neural networks, where data is encoded into the parametrized Hamiltonian of a quantum system and evolves under quantum dynamics \cite{NMGPSZ:21:CommPhys,ILPFFP:23:CommPhys}. Observables, such as Pauli operators, extract transformed data representations, which are then processed using a classical trainable readout layer. Conceptually, this approach bypasses the optimization bottlenecks of VQCs and avoids barren plateaus. Beyond efficiency, QRC leverages intrinsic physical processes, such as noise and dissipation, as computational resources, thereby enhancing temporal memory and forecasting while preserving universality in approximating fading memory maps. Noise-mapping techniques also enable the precise characterization of circuit dynamics under realistic hardware conditions, ensuring that QRC implementations remain effective on near-term devices \cite{SMSGZ:24:Quantum,G:24:arXiv}. Crucially, the parameters that govern the non-linear interactions and dynamical regime of the reservoir remain fixed during training, bypassing the limitations inherent in variational approaches. This approach excels in dynamical-system-level expressiveness, particularly for tasks based on temporal and sequential data such as time-series prediction, forecasting, and anomaly detection. Recent large-scale implementations on hardware architectures, including Gaussian boson sampling and neutral atom-based quantum computing platforms, have demonstrated the broad applicability and competitive performance of QRC across machine learning tasks \cite{KHZWWHZCHZBAO:24:arXiv, CSPMMYWOWM:25:arXiv}.

As AI becomes integral to critical safety applications, from healthcare and autonomous driving to security systems, ensuring robustness against adversarial attacks is paramount \cite{YBK:18:NBE, MULLDA:20:IEEE_J_ITS, SFN:21:SNCS}. In classical machine learning, adversarial examples reliably expose vulnerabilities, sparking extensive defense research \cite{KGB:16:arXiv}, Parallel studies show that quantum classifiers, particularly VQC-based models, also exhibit adversarial fragility, with required perturbation strengths decreasing as system size grows \cite{WELSHU:23:PRR, LDD:20:PRR, WTLHHHEU:23:Nat_Mach_Intel}. These vulnerabilities persist under both white-box and black-box adversarial attacks, with quantum noise and decoherence offering only limited protection \cite{WTD:24:IEEE_QCE,YRKVVWC:24:arXiv}. Furthermore, recent benchmark studies have revealed asymmetric attack transferability, where perturbations crafted on quantum models tend to fool classical networks more effectively than the reverse \cite{WECLSHMU:23:PRR, EMSSB:24:IEEE_ICIPCW}.

Parallel studies have explored defense strategies such as randomized quantum encodings to suppress adversarial gradients~\cite{GYLD:24:PRR} and data augmentation to harden quantum kernel methods against input perturbations~\cite{MB:25:QMI}. However, while the adversarial robustness of QML classifiers has been extensively investigated, the robustness characteristics of QRC models—particularly those employing analog Rydberg-atom implementations—remain largely unexplored. Here, we present the first systematic study of adversarial robustness in a Rydberg-based QRC framework. Specifically, we examine whether augmenting lightweight MLPs with quantum reservoir embeddings enhances resilience against gradient-based perturbations. Under a white-box threat model, where the adversary has full access to model parameters and gradients, robustness is evaluated using three canonical attacks: the \ac{FGSM}~\cite{EMMSSB:24:ICIPCW,GSS:15:arXiv,MMSTV:19:arXiv}, \ac{PGD}~\cite{YRKVWC:24:arXiv,MMSTV:19:arXiv,WNHCHSU:24:IC,MCKKAFNCM:25:arXiv}, and DeepFool~\cite{WELSHU:23:PRR,MFF:16:CVPR,TPO:22:WACV}. Empirical evaluations across the \textsc{MNIST}, \textsc{Fashion-MNIST}, and \textsc{Kuzushiji-MNIST} benchmarks demonstrate that integrating a Rydberg quantum reservoir with a classical readout consistently enhances adversarial robustness, highlighting a scalable and hardware-realistic pathway for robust quantum learning.

In this context, it is increasingly important to emphasize practical robustness and hardware realizability rather than purely asymptotic arguments. Reservoir-style models offer low-overhead readout and intrinsic echo-state properties that remain stable even under noise, highlighting their suitability for near-term devices. At the same time, theoretical perspectives caution that avoiding issues such as the curse of dimensionality or barren plateaus may confine models to effective subspaces that are classically simulable \cite{CLGDBF:25:Nat_Comm,JWZZLC:25:arXiv,DLCLGLPR:25:arXiv}. These insights strengthen the motivation to frame QRC not around abstract speedups, but as a pathway to robust, tunable, and hardware-realistic models—qualities that our Rydberg-based Hamiltonian implementation seeks to exploit in adversarial settings.

This work addresses this gap by investigating the adversarial robustness of QRC implemented via Rydberg Hamiltonians. We evaluate whether augmenting multilayer perceptrons \ac{MLP} with quantum reservoir embeddings enhances resilience to adversarial perturbations, providing the first systematic study of QRC robustness in adversarial settings \cite{GYLD:24:PRR,MB:25:QMI,EMSSB:24:IEEE_ICIPCW}. Our findings suggest that Rydberg-based QRC provides a hardware-realistic and robust pathway for QML, advancing both theoretical understanding and practical deployment in adversarial settings.

\begin{figure}[!t]
    \centering
    \begin{subfigure}[b]{\linewidth}
        \centering
        \includegraphics[width=1\linewidth]{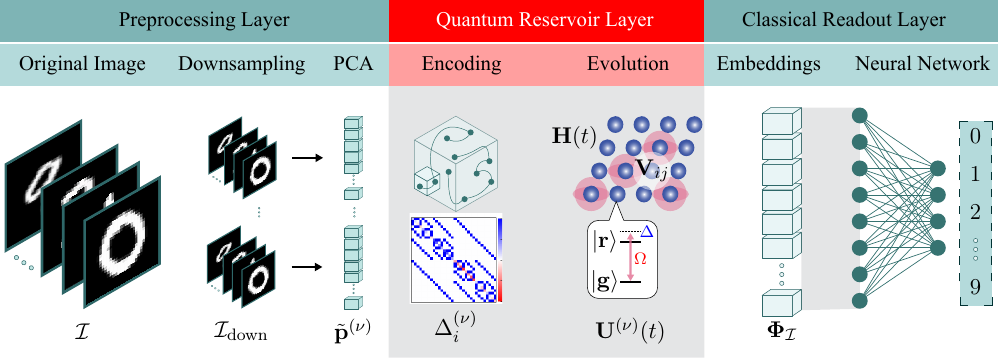}
        \caption{}
        \label{fig:1a}
    \end{subfigure}

    \vspace{1em}

    \begin{subfigure}[b]{\linewidth}
        \centering
        \includegraphics[width=0.98\linewidth]{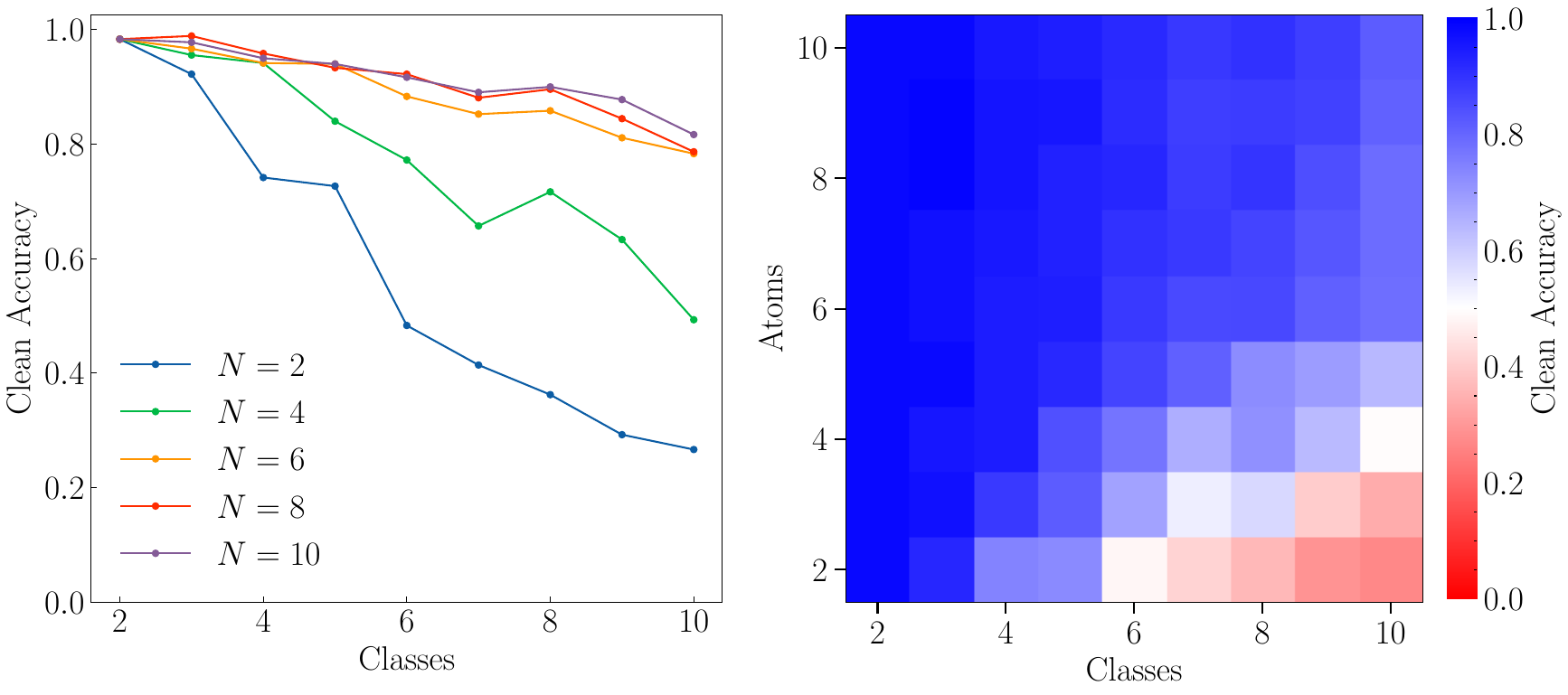}
        \caption{}
        \label{fig:clean-acc}
    \end{subfigure}
    \caption{
    (a) QRC based learning Framework.
    (b) Classification performance and robustness of the quantum reservoir on a balanced MNIST subset. Here, we use a balanced MNIST dataset comprising $100$ handwritten-digit samples per class ($10$ classes in total), with a $70\%$/$30\%$ train-test split. We keep $\delta = N $ here.  (Left) Clean-test accuracies achieved by a fixed classical readout layer when driven by quantum reservoirs under different configurations (atoms) for $N$. (Right) Dependence of classification robustness on the dimensionality $N$ of the reservoir’s learning space.
    }
    \label{fig:fig1_combined}
\end{figure}

\begin{figure}[!t]
    \centering

        \centering
        \includegraphics[width=1\linewidth]{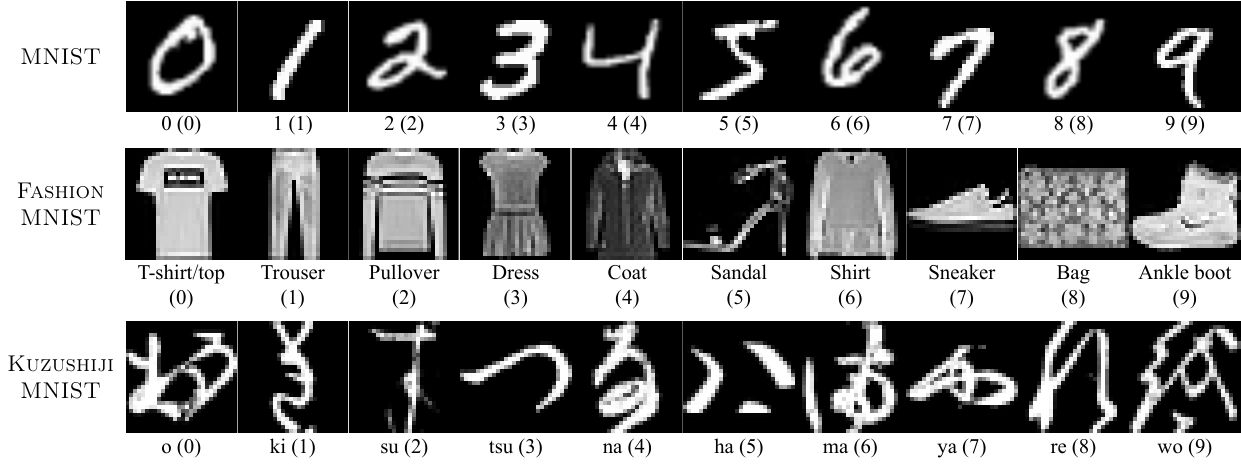}
        \caption{Sample images from the three benchmark datasets employed in this study: MNIST, Fashion-MNIST, and Kuzushiji-MNIST. Each dataset consists of 10 balanced classes used to rigorously evaluate the adversarial robustness of the quantum reservoir learning model.}
        \label{fig:2}

\end{figure}

\begin{figure}[!t]
  \centering
  \begin{subfigure}[t]{0.20\linewidth}
    \centering
    \includegraphics[width=\linewidth]{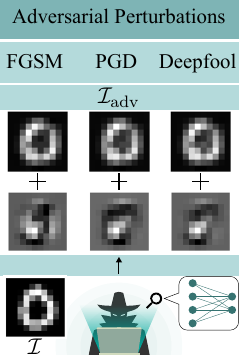}
    \label{fig:adv-acc-left}
  \end{subfigure}
  \hfill
  \begin{subfigure}[t]{0.79\linewidth}
    \centering
\includegraphics[width=\linewidth,height=0.195\textheight]{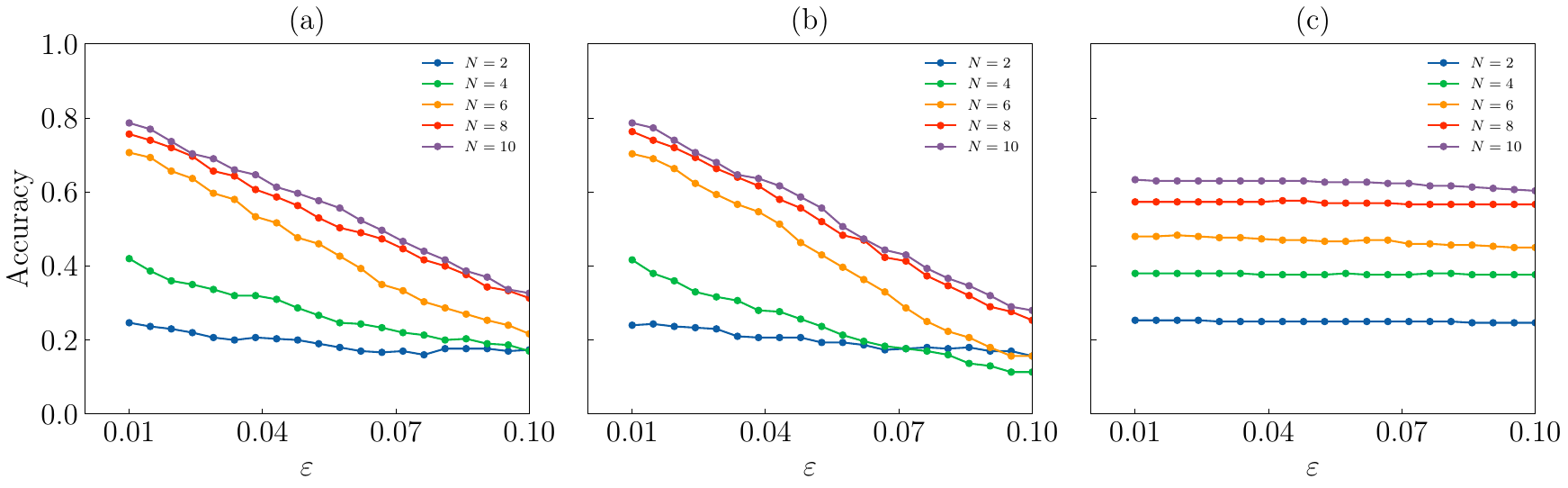}
  \end{subfigure}

    \caption{Classification performance and adversarial robustness of the quantum reservoir on a balanced MNIST subset of all the $10$ classes under three attacks. (a) FGSM (b) PGD (c) Deepfool. For all the attacks, we vary the budget $\varepsilon \in [0.0, 0.1]$, with $100$ gradient steps and perturbation rate of $10^{-3}.$ }
  \label{fig:adv-acc}
\end{figure}

\section*{Results}\label{res}

\subsection*{Simulation Setup}
The dynamics of the quantum reservoir-based learning are simulated using NVIDIA's CUDA-Q platform~\cite{BCCCFGGGGH:23:QCE23}, which utilizes \ac{GPU} acceleration for efficient quantum evolution of many-body systems through built-in numerical integrators. The Rydberg atom array is configured with experimentally validated parameters from ~\cite{KHZWWHZCHZBAO:24:arXiv} as shown in Table~\ref{tab:params}, where we choose a uniform modulation $\alpha_i = 0.15$ for all sites, effectively controlling the strength of the local detuning relative to the global drive. Moreover, the time-dependent Hamiltonian was constructed as a sparse operator with CUDA-optimized evaluation of interaction terms,$\B{V}_{ij}$, parallelized over the GPU threads. The embedding vector $\B{\Phi}({x}^{(k)})\in\mathbb{R}^D$ is passed into a three‐layer MLP readout. The network has an input dimension $D$, a first hidden layer of size $64$, a second hidden layer of size $32$, and an output layer of size $C$, which equals the number of classes. Each hidden layer applies a ReLU activation followed by a dropout rate of $10^{-3}$. We train only these MLP parameters using the Adam optimizer with learning rate $10^{-3}$, batch size $64$, and for up to $500$ epochs, minimizing the softmax cross‐entropy loss. Since the quantum reservoir itself remains fixed, all learnable parameters reside in this lightweight classical readout, keeping training efficient. In this study, we use the balanced \textsc{MNIST}, \textsc{Fashion-MNIST}, and \textsc{Kuzushiji-MNIST} datasets comprising $100$ samples per class ($700$ samples for $10$ classes total), with a $70\%/30\%$ train-test split ratio.


\begin{table}[!htbp]
\centering
\caption{Physical parameters for the Rydberg reservoir}
\label{tab:params}
\begin{tabular}{@{}lccc@{}}
\toprule
\textbf{Parameter} & \textbf{Symbol} & \textbf{Value} & \textbf{Unit} \\
\midrule
Number of atoms & $N$ & 8 & -- \\
Lattice spacing & $d$ & 10.0 & $\umu$m \\
Interaction coefficient & $C_6$ & $2\pi \times 2000$ & MHz$\cdot\umu\mathrm{m}^6$ \\
Rabi frequency & $\Omega$ & $2\pi \times 5$ & MHz \\
Detuning range & $[\Delta_{\min}, \Delta_{\max}]$ & $[0, 2\pi \times 10]$ & MHz \\
Local detuning modulation & $\alpha$ & 0.15 & -- \\
Total evolution time & $T$ & 3.0 & $\umu$s \\
Time steps & $M$ & 6 & -- \\
Initial state & $\ket{\psi_0}$ & $|+\rangle^{\otimes N}$ & -- \\
\bottomrule
\end{tabular}
\end{table}


\subsection*{Encoding Images}
Classical image data requires specialized encoding to interface with quantum reservoirs due to dimensionality mismatches between pixel spaces and qubit resources. In this paper, we employ a multi-stage approach:

\subsubsection*{Image Preprocessing}
Original images $\mathcal{I} \in \mathbb{R}^{L \times L}$ are first downsampled to $\mathcal{I}_{\text{down}} \in \mathbb{R}^{S \times S}$ using area interpolation (OpenCV) or Lanczos resampling (PIL). This reduces computational complexity and filters high-frequency noise while preserving structural information:
\begin{align}
\mathcal{I}_{\text{down}} = \mathrm{resize}\bigl(\mathcal{I},(S,S)\bigr)
\end{align}
Each downsampled image is decomposed into $\kappa$ non-overlapping patches $\{\B{p}^{(\nu)}\}_{\nu=1}^{\kappa}$ with $\B{p}^{(\nu)} \in \mathbb{R}^{P^{2}}$, where $P$ is the patch width; hence $\kappa = (S/P)^{2}$. Patch extraction enables localized feature analysis, capturing spatial redundancies inherent in natural images:
\begin{align}
\B{p}^{(\nu)} = \mathrm{extract}_{\nu}\bigl(\mathcal{I}_{\text{down}}\bigr)
\end{align}
Each patch is projected to $\tilde{\B{p}}^{(\nu)}\in\mathbb{R}^{\delta}$, with $\delta\le N$ matching the number of atoms. Dimensionality reduction via \ac{PCA} ($\B{W}$) decorrelates patch features and compresses data by retaining maximal variance directions :
\begin{align}
\tilde{\B{p}}^{(\nu)} = \B{W}^{\top}\!\bigl(\B{p}^{(\nu)} - \B{\mu}\bigr)
\end{align}
Here $\B{W}\in\mathbb{R}^{P^{2}\times\delta}$ contains the principal components and $\B{\mu}$ is the global patch mean.  The projection matrix solves the eigenproblem
\begin{align}
\B{\B{\sigma}}\,\B{W} = \B{W}\,\B{\Lambda},\qquad
\B{\B{\sigma}} = \frac{1}{\kappa} \sum_{\nu=1}^{\kappa} \bigl(\B{p}^{(\nu)}-\B{\mu}\bigr)\bigl(\B{p}^{(\nu)}-\B{\mu}\bigr)^{\!\top}
\end{align}
where $\B{\Lambda}=\diag(\lambda_{1},\dots,\lambda_{\delta})$ contains the largest $\delta$ eigenvalues chosen such that
$\frac{\sum_{i=1}^{\delta}\lambda_{i}}{\sum_{i=1}^{P^{2}}\lambda_{i}} \;>\; \mu,$ thus preserves at least a fraction $\mu$ of the total variance. This variance-retention criterion, standard in dimensionality reduction, ensures that the compressed representation remains information-rich while discarding redundant components.

\begin{figure}[t]
    \centering
    \includegraphics[width=\linewidth]{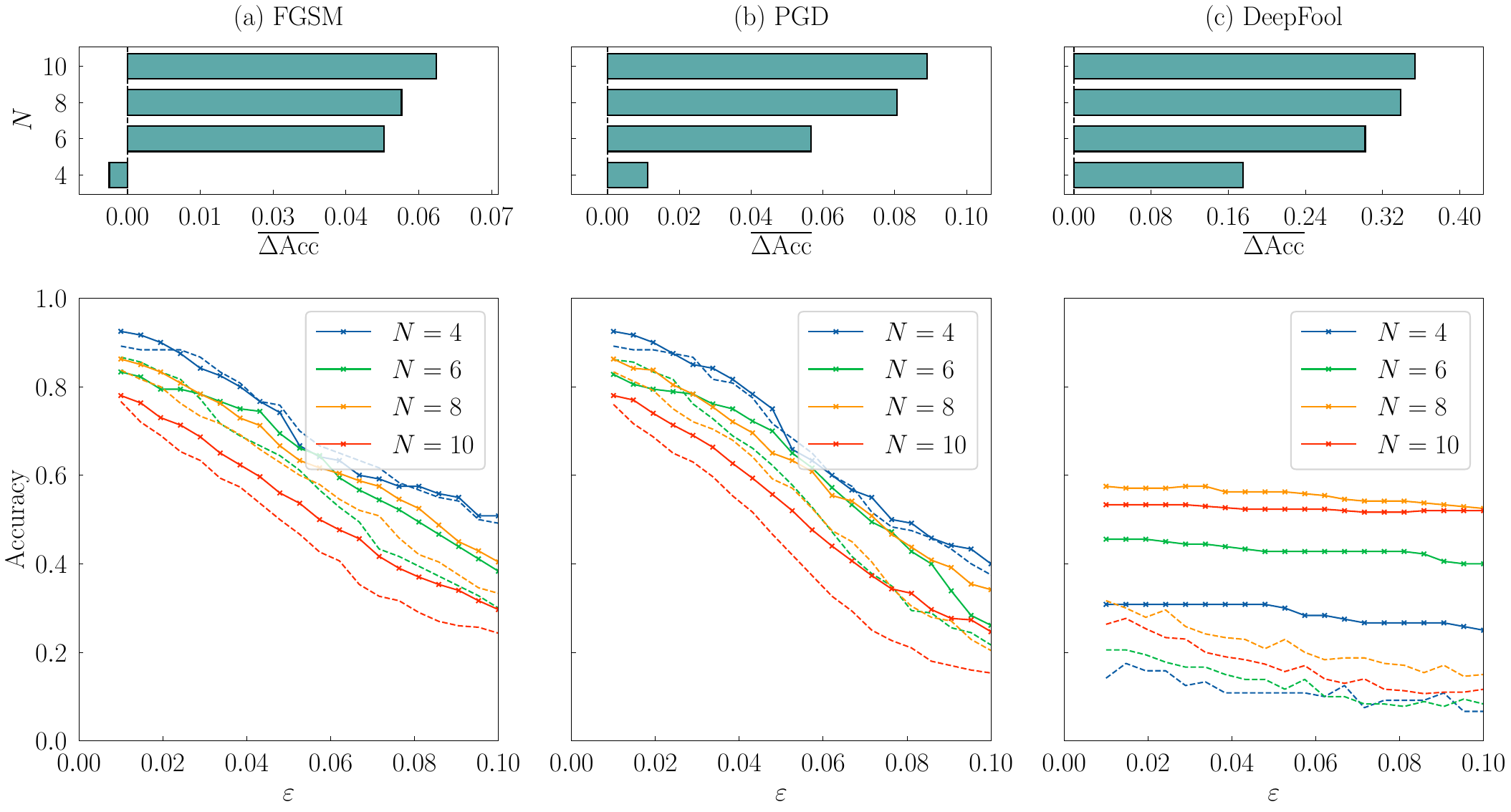}
    \caption[\textsc{MNIST} adversarial robustness]{\CommonAdvCaption\ \textbf{Dataset:} \textsc{MNIST}.}
    \label{fig:adv-all-mnist}
\end{figure}

\begin{figure}[t]
    \centering
    \includegraphics[width=\linewidth]{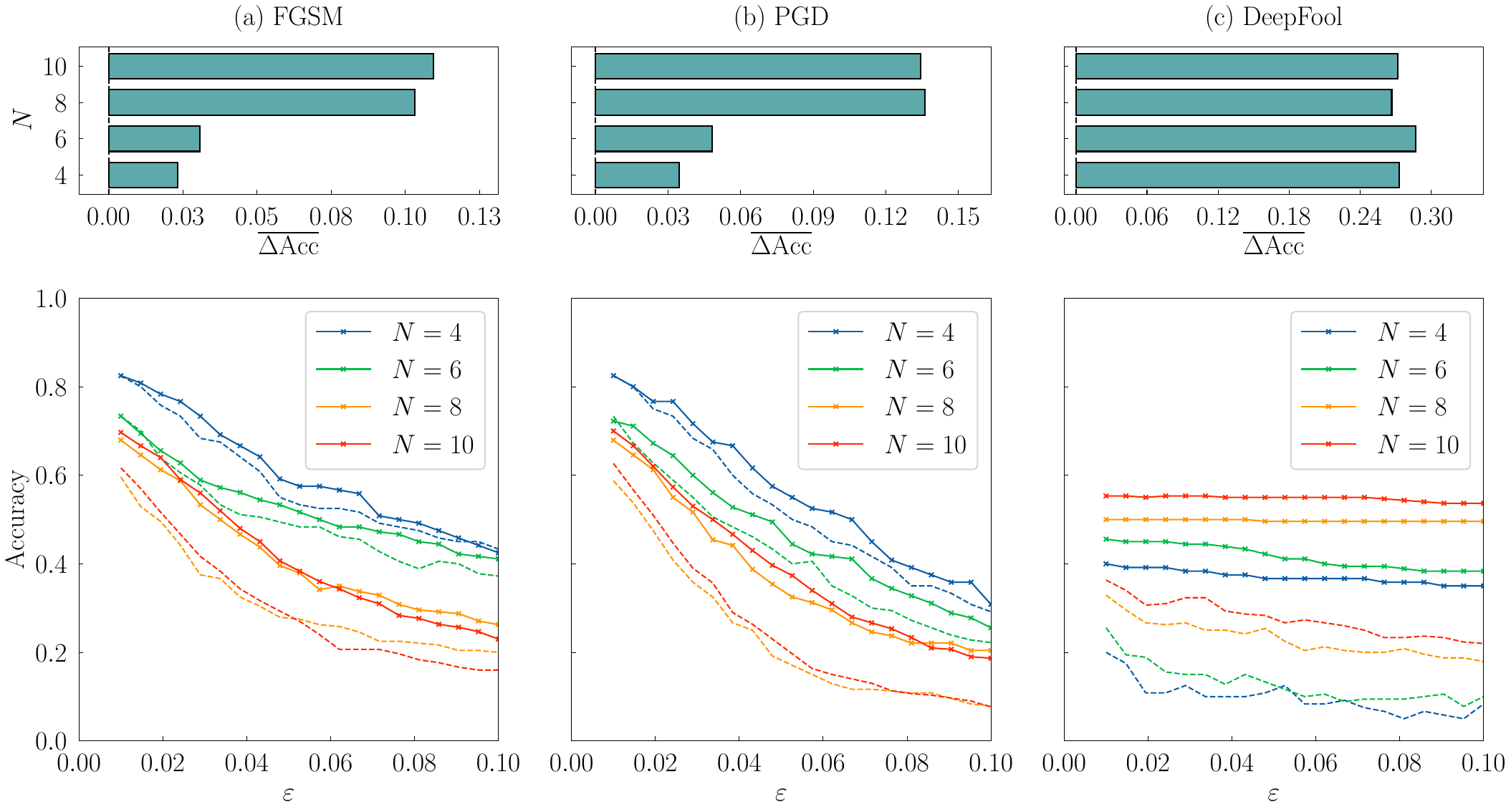}
    \caption[\textsc{Fashion\textnormal{-}MNIST} adversarial robustness]{\CommonAdvCaption\ \textbf{Dataset:} \textsc{Fashion-MNIST}.}
    \label{fig:adv-all-fmnist}
\end{figure}

\begin{figure}[t]
    \centering
    \includegraphics[width=\linewidth]{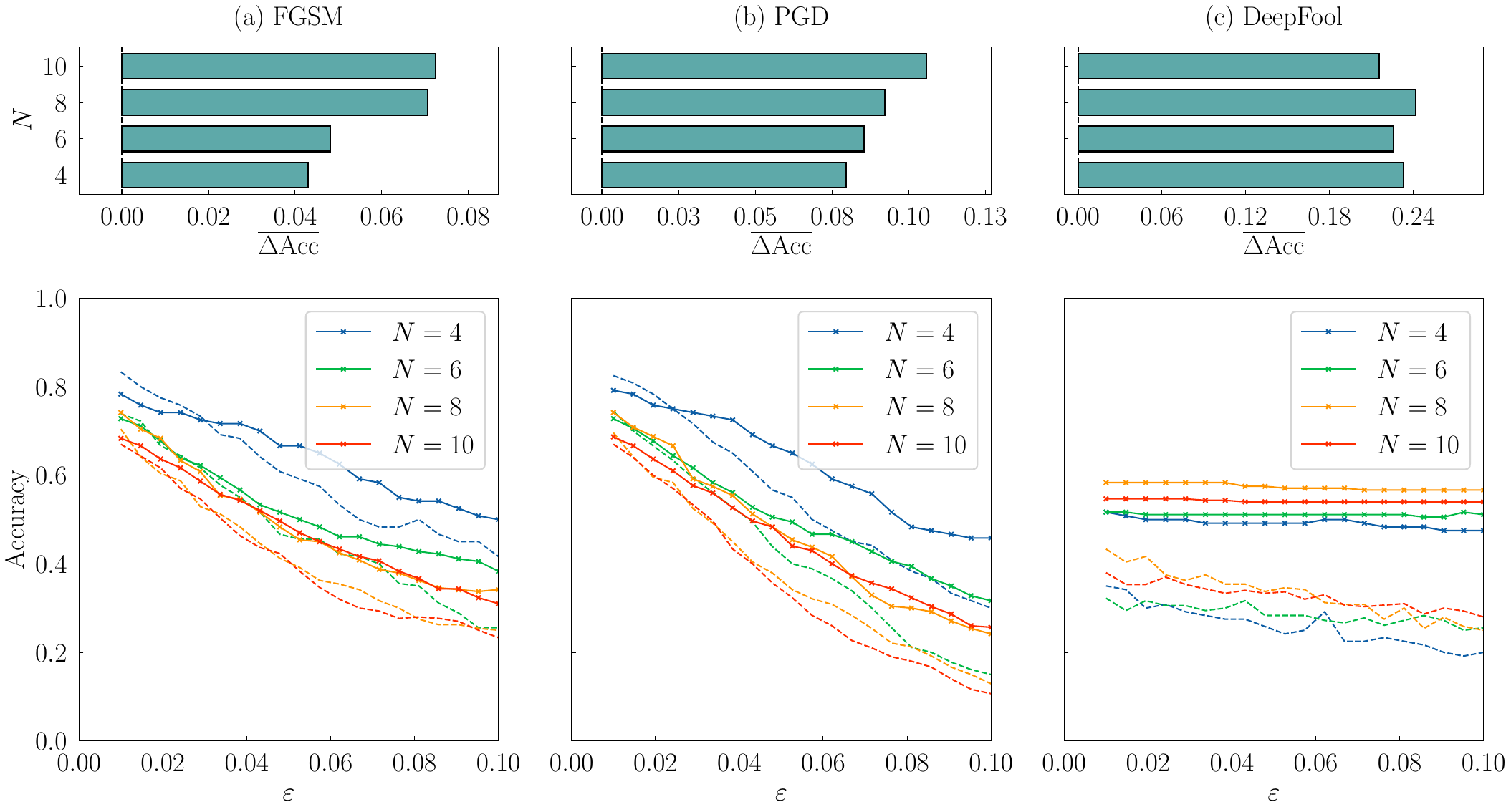}
    \caption[\textsc{Kuzushiji\textnormal{-}MNIST} adversarial robustness]{\CommonAdvCaption\ \textbf{Dataset:} \textsc{Kuzushiji-MNIST}.}
    \label{fig:adv-all-kmnist}
\end{figure}

\subsubsection*{Quantum Encoding}
Compressed features are mapped to detunings:
\begin{align}
\Delta_i^{(\nu)} &= \Delta_{\min} + \alpha_i^{(\nu)}\bigl(\Delta_{\max}-\Delta_{\min}\bigr),\\[2pt]
\alpha_i^{(\nu)} &= 
\frac{\tilde{p}_i^{(\nu)}-\displaystyle\min_{\nu'}\tilde{p}_i^{(\nu')}}%
     {\displaystyle\max_{\nu'}\tilde{p}_i^{(\nu')}-\displaystyle\min_{\nu'}\tilde{p}_i^{(\nu')}}.
\end{align}
The Hamiltonian for patch $\nu$ is
\begin{align}
\B{H}^{(\nu)}(t)=
\sum_{i=1}^{N}\frac{\Omega(t)}{2}\,\B{\sigma}_i^{x}
-\sum_{i=1}^{N}\Delta_i^{(\nu)}\,\B{\hat{n}}_i
+\sum_{i<j} \B{V}_{i j}\,\B{\hat{n}}_i\B{\hat{n}}_j,
\end{align}
where $\B{\hat{n}}_i=\tfrac12(\B{I}-\B{\sigma}_i^{z})$.Each patch undergoes independent reservoir evolution. Observables are sampled at $M$ time points $\{t_m\}_{m=1}^{M}$, yielding
\begin{align}
\B{\Phi}^{(\nu)}(t_m)=
\Bigl(
\langle\B{\sigma}_1^{z}\rangle_{t_m}^{(\nu)},\dots,\langle\B{\sigma}_{N}^{z}\rangle_{t_m}^{(\nu)},
\langle\B{\sigma}_1^{z}\B{\sigma}_2^{z}\rangle_{t_m}^{(\nu)},\dots,
\langle\B{\sigma}_{N-1}^{z}\B{\sigma}_{N}^{z}\rangle_{t_m}^{(\nu)}
\Bigr)^{\!\top}.
\end{align}
Concatenating all $M$ snapshots gives the patch embedding
\begin{align}
\B{\Phi}^{(\nu)}=
\bigl(
\B{\Phi}^{(\nu)}(t_1)^{\top}\!
\mid\!
\cdots\!
\mid\!
\B{\Phi}^{(\nu)}(t_M)^{\top}
\bigr)^{\!\top}\in\mathbb{R}^{\phi},
\quad
\phi=M\Bigl(N+\tbinom{N}{2}\Bigr).
\end{align}
The image-level representation is obtained by patch averaging:
\begin{align}
\B{\Phi}_{\mathcal{I}}=\frac{1}{\kappa}\sum_{\nu=1}^{\kappa}\B{\Phi}^{(\nu)}.
\end{align}
This hierarchical encoding leverages quantum dynamics for intra-patch feature extraction, while utilizing classical aggregation for inter-patch dimensionality reduction.

\subsection*{Adversarial Attacks}
Adversarial attacks craft worst‑case inputs that force a trained model to
misclassify.  In this paper, we focus on three canonical white-box attacks, namely, FGSM, PGD, and DeepFool, that assume full access to network gradients.  Each method returns an adversarial image $\mathcal{I}_{\mathrm{adv}}$ whose perceptual distance from the original image $\mathcal{I}$ does not exceed a chosen attack \emph{budget}~$\varepsilon$. In this paper, we vary the $\varepsilon \in [0.0, 0.1]$, with $100$ gradient steps and a learning rate of $10^{-3}$.

\subsubsection*{FGSM}
FGSM is a foundational adversarial technique that generates perturbed inputs by leveraging the direction of steepest ascent in the model’s loss landscape. It operates by applying a single targeted distortion to the input in the direction that most increases prediction error, thus inducing misclassification with minimal computational effort. This method exploits local linearity in neural networks and serves as an efficient diagnostic of model vulnerability. In the context of this study, FGSM serves as a benchmark for baseline robustness, where quantum reservoir models consistently enhance their classical counterparts by resisting shallow gradient-based manipulations, indicating a fundamental mismatch between classical perturbation geometry and quantum feature encoding.
FGSM linearises the loss $\mathcal{L}(\mathcal{I},y)$ around the clean image and
moves $\varepsilon$ in the direction that maximally increases the loss in
$\ell_{\infty}$ geometry:
\begin{align}
  \mathcal{I}_{\mathrm{adv}}
  &= \mathcal{I}
     + \varepsilon\,
       \mathrm{sign}\bigl(\nabla_{\mathcal{I}}\mathcal{L}(\mathcal{I},y)\bigr),
  \qquad
  \lVert\mathcal{I}_{\mathrm{adv}}-\mathcal{I}\rVert_{\infty}\le\varepsilon .
\end{align}
A single forward–backward pass suffices, making FGSM a lightweight
\textquote{stress test}.  Budgets in image classification typically span
$\varepsilon\!\in\!\{8/255,\,16/255\}$ for inputs scaled to~$[0,1]$.

\subsubsection*{PGD}

PGD extends the principle of FGSM into a more aggressive and iterative regime. It applies repeated, controlled perturbations that stay within a bounded region, systematically exploring the model’s decision boundaries to expose deeper vulnerabilities. Each iteration applies a limited displacement in the gradient direction, maintaining control over the perturbation’s size at every step. It is widely regarded as a rigorous adversarial test due to its ability to converge on high-impact inputs. Within this study, PGD is used to assess the stability of quantum and classical models under sustained adversarial perturbation. PGD refines FGSM into an iterative constrained optimization that maximizes the loss over the $\ell_{\infty}$ ball of budget~$\varepsilon$. From an initial point $\mathcal{I}_{0}$ (often random inside the ball), it
performs $T$ steps of size~$\zeta$:
\begin{align}
  \mathcal{I}_{t+1} \;=\;
  \operatorname{Proj}_{\lVert\boldsymbol{\eta}\rVert_{\infty}\le\varepsilon}
  \!\Bigl(
     \mathcal{I}_{t}
     + \zeta\,\mathrm{sign}\bigl(\nabla_{\mathcal{I}_{t}}
       \mathcal{L}(\mathcal{I}_{t},y)\bigr)
  \Bigr),
  \qquad t=0,\dots,T-1.
\end{align}
For step sizes $\zeta\!\approx\!2/255$ and $T\!\approx\!40$, PGD closely
approximates the worst‑case $\ell_{\infty}$ perturbation and is the de‑facto
benchmark for robustness studies.

\subsubsection*{DeepFool} 

DeepFool is a geometry-driven attack that seeks to identify the closest point at which an input crosses the decision boundary of a classifier, producing adversarial examples with minimal perceptual distortion. It operates by incrementally adjusting the input until it reaches the region of misclassification, effectively modeling the local topology of the decision boundary. This attack provides insight into the structural vulnerability of models. More specifically, it seeks the smallest $\ell_{2}$ perturbation that crosses the current decision boundary.  At each iteration, the classifier is locally linearized into a hyperplane, and the input is nudged orthogonally toward it:
\begin{align}
  \boldsymbol{\eta}^{\star}
  &= -\frac{f(\mathcal{I})}
         {\lVert\nabla f(\mathcal{I})\rVert_{2}^{2}}\,
       \nabla f(\mathcal{I}),
  \qquad
  \mathcal{I}\;\leftarrow\;\mathcal{I} + \boldsymbol{\eta}^{\star},
\end{align}
where $f$ is the signed decision function.  Iterations stop once the predicted
label changes.  To enforce the attack budget~$\varepsilon$, the final perturbation
is rescaled if necessary:
\begin{align}
  \boldsymbol{\eta}
  = \min\!\Bigl\{1,\tfrac{\varepsilon}{\lVert\boldsymbol{\eta}^{\star}\rVert_{2}}
    \Bigr\}\,\boldsymbol{\eta}^{\star},
  \qquad
  \lVert\boldsymbol{\eta}\rVert_{2}\le\varepsilon,\quad
  \mathcal{I}_{\mathrm{adv}}=\mathcal{I}+\boldsymbol{\eta}.
\end{align}
DeepFool typically converges in fewer than ten iterations and yields
quasi‑imperceptible $\ell_{2}$ perturbations.

\section*{Discussion}\label{conclusion}
This work provides a systematic investigation of adversarial robustness in quantum reservoir learning. The results show that augmenting a classical MLP with a Rydberg-based quantum reservoir improves both clean and adversarial accuracies. By demonstrating enhanced performance without requiring variational training, the proposed framework establishes Rydberg reservoirs as practical and scalable components for robust QML on near-term quantum processors. In particular, we investigated the robustness of the proposed QRC--MLP architecture across varying reservoir dimensions $N$. For each configuration, PCA was applied to project input images into $\delta = N$ principal components, ensuring feature--reservoir dimensional consistency. As shown in Fig.~\ref{fig:clean-acc}, larger reservoirs consistently achieved higher clean accuracies, indicating that the Rydberg-based reservoir enriches the nonlinear feature space accessible to the classical readout. Under the \ac{CD} encoding scheme---where each feature corresponds to one atom---the condition $N \ge \delta$ ensures sufficient expressiveness and provides a clear baseline for assessing scalability.

Adversarial evaluations under FGSM, PGD, and DeepFool attacks further confirm the benefit of reservoir augmentation. In Figs.~\ref{fig:adv-all-mnist}--\ref{fig:adv-all-kmnist}, the hybrid QRC--MLP model maintains higher accuracy than the purely classical MLP across all perturbation budgets. The robustness improvement, expressed as the mean accuracy enhancement
\begin{align}
\overline{\Delta\mathrm{Acc}}_{N}
=\frac{1}{|\mathcal{E}|}\sum_{\varepsilon\in\mathcal{E}}
\Big(\mathrm{Acc}^{\text{QRC}+\text{MLP}}_{N}(\varepsilon)
-\mathrm{Acc}^{\text{MLP}}_{N}(\varepsilon)\Big),
\\
\mathrm{Acc}^{(\cdot)}_{N}(\varepsilon)
=\frac{1}{|\mathcal{D}|}\sum_{(x,y)\in\mathcal{D}}
\mathbb{1}\!\left\{\hat{y}^{(\cdot)}_{N,\varepsilon}(x)=y\right\},
\end{align}
increases with $N$, demonstrating that the high-dimensional embeddings generated by the reservoir are less susceptible to gradient-based perturbations. Across all attack methods, configurations with $N \ge \delta$ sustain higher accuracies throughout the entire perturbation range $\varepsilon \in [0,0.1]$, confirming that Rydberg interactions play a decisive role in shaping adversarial robustness.

Although reservoir augmentation provides a clear advantage, the overall robustness depends strongly on the design of the quantum reservoir. The initialization state~$\psi_0$, the choice of observables~$\mathbf{\hat{O}}$, and the encoding function that maps classical data to quantum detunings collectively determine the expressive capacity of the system. These factors directly influence how information is distributed within the reservoir and how effectively temporal correlations are captured. Moreover, realistic evaluations must include the impact of decoherence, parameter drift, and cross-talk in actual Rydberg hardware. Incorporating these noise effects in future studies will be essential to accurately reflect performance on near-term devices.

This work also highlights that increasing the reservoir dimension~$N$ enhances robustness but introduces practical trade-offs in qubit resources and measurement overhead. Despite these challenges, the proposed Rydberg-based QRC architecture remains attractive because it achieves robustness without requiring gradient-based optimization. This property makes it especially suited for \ac{NISQ} systems, where stability and resource efficiency are critical for implementation feasibility.

In future studies we aim to extend the current framework toward more diverse data modalities, such as spatiotemporal data and time series, while exploring systematic methods for tuning Hamiltonian parameters to further improve performance. Additionally, selecting appropriate observables and refining encoding strategies will help maximize information extraction and minimize redundancy. Experimental validation on real quantum hardware, such as neutral-atom or superconducting platforms under realistic noise and sampling constraints, will be a crucial next step in verifying the robustness advantages observed in simulation. Moreover, incorporating realistic noise models and extending the reservoir-augmentation framework to advanced architectures such as transformers will be vital for translating these quantum advantages into practical, robust machine learning systems.

\section*{Methods}\label{Methods}

\subsection*{Quantum Reservoir Computing}\label{QRL}
QRC builds upon the principles of classical reservoir computing by exploiting the high-dimensional dynamics and quantum parallelism inherent in quantum systems for spatiotemporal representation learning~\cite{ZEN:24:arXiv, JLHKSCWHH:24:Adv_Mater}. The QRC based learning framework is illustrated in Fig.~\ref{fig:1a}. Unlike quantum neural networks, which require extensive training across multiple layers of adjustable parameters, QRC utilizes a fixed, untrained quantum reservoir. The reservoir functions as a natural feature map, where the input data is encoded in the Hamiltonian of the system, and the resulting quantum dynamics transforms the input into a non-linear high-dimensional representation~\cite{KHZWWHZCHZBAO:24:arXiv}. 
QRC requires no iterative training within the reservoir, restricting the learning complexity to optimizing only the readout layer and greatly simplifying implementation on NISQ devices, where resource constraints and noise hinder large-scale parameter tuning. Adaptation to a specific learning task is effected through an offline configuration of the reservoir’s dynamics to tune Hamiltonian parameters by using genetic or other meta‐optimization algorithms\cite{XZQCZLDL:23:Sci_Bull}.This approach is particularly well-suited for extracting temporal patterns and sequential data, akin to classical \ac{RNN}-based echo-state networks or \ac{LSTM}. Various physical implementations have been explored, including coherently coupled quantum oscillators~\cite{DCPMGM:23:npj_QI}, Rydberg atom arrays or neutral-atom quantum computing systems~\cite{BNGY:22:PRXQ}, superconducting quantum devices~\cite{YSKNGZSNYY:23:arXiv}, and Gaussian boson samplers~\cite{CSPMMYWOWM:25:arXiv}. In general, the system Hamiltonian is modeled as
\begin{align}
\B{H} = \B{H}_0 + \B{H}_{\text{int}} + \B{H}_{\text{drive}}(t),
\end{align}
where $\B{H}_0$ represents the internal energy, $\B{H}_{\text{int}}$ encodes interactions between constituents, and $\B{H}_{\text{drive}}(t)$ incorporates the data as time-dependent driving fields. The temporal evolution of the reservoir provides a rich, non-linear mapping from the input data space to a high-dimensional Hilbert space. The classical features for downstream tasks are extracted through local observables (e.g., Pauli operators). The reservoir Hamiltonian is initially designed to meet the constraints of quantum hardware, such as the neural atom-based quantum computer \cite{KHZWWHZCHZBAO:24:arXiv}, and can be optionally tuned with offline meta-optimization; Once configured, it remains unchanged, and all subsequent batched or online learning is applied exclusively to the classical readout layer, preserving the original dynamics of the reservoir.
In the subsequent section, we model the end-to-end framework used in this study, based on the Rydberg Hamiltonian described in~\cite{KHZWWHZCHZBAO:24:arXiv}.

\subsubsection*{Rydberg Hamiltonian}
Consider a system of $N$ neutral atoms arranged in a one-dimensional lattice, each modeled as a two-level system with ground state $\ket{g}$ and Rydberg excited state $\ket{r}$. The many-body dynamics under Rydberg blockade are governed by the time-dependent Hamiltonian (with $\hbar = 1$):
\begin{align}
\B{H}^{(k)}(t)
&= \sum_{i=1}^{N} \frac{\Omega(t)}{2}\,\B{\sigma}_i^x
\;-\;\sum_{i=1}^{N} \alpha_i\,\Delta_i^{(k)}\,\B{\hat{n}}_i
\;+\;\sum_{i<j} \B{V}_{ij}\,\B{\hat{n}}_i\B{\hat{n}}_j,
\end{align}

where $\B{\sigma}_i^x = \ket{g_i}\!\bra{r_i} + \ket{r_i}\!\bra{g_i}$ is the Pauli-$x$ operator for atom $i$, and $\B{\hat{n}}_i = \tfrac{1}{2}(\B{I}-\B{\sigma}_i^z)$ is the projector onto the Rydberg state. Here, $\alpha_i\in[0,1]$ is a site-dependent modulation factor that scales the feature-encoded detuning $\Delta_i^{(k)}$ at atom $i$. The interaction coefficients $\B{V}_{ij} = C_6/|\B{r}_i - \B{r}_j|^6$ represent van der Waals interactions between atoms $i$ and $j$. Given a data set of $n$ samples with feature vectors $\B{x}^{(k)} = \bigl(x_1^{(k)},\,x_2^{(k)},\,\dots,\,x_N^{(k)}\bigr)^{\!\mathsf T} \in \mathbb{R}^N$, we employ CD encoding by mapping each feature to a static detuning via min–max normalization:
\begin{align}
\Delta_i^{(k)} = \Delta_{\min} + \left(\frac{x_i^{(k)} - x_{\min}}{x_{\max} - x_{\min}}\right)\bigl(\Delta_{\max} - \Delta_{\min}\bigr),
\end{align}
where $x_{\min} = \min_{i,k} x_i^{(k)}$ and $x_{\max} = \max_{i,k} x_i^{(k)}$ are computed across the data set. The range $[\Delta_{\min}, \Delta_{\max}]$ defines the bounds of the applied detunings. This encoding results in sample-specific, time-independent Hamiltonians where the input features are embedded as local energy shifts. The Rabi frequency $\Omega(t)$ quantifies the rate of coherent population oscillations between the ground state $\ket{g}$ and the Rydberg state $\ket{r}$, which control the quantum dynamics of the reservoir through external laser fields that set both the global drive strength and the site-dependent detunings. Both the Rabi frequency $\Omega(t)$ and the site-dependent detunings $\Delta_i^{(k)}$ are controlled by laser parameters: $\Omega(t)$ corresponds to the time-dependent laser amplitude driving coherent Rabi oscillations, while each $\Delta_i^{(k)}$ corresponds to the local detuning of the laser from resonance at atom $i$. The drive $\Omega(t)$ is applied globally and kept fixed across samples, whereas the detunings $\Delta_i^{(k)}$ encode input-dependent information and remain static during evolution.

\subsubsection*{Constructing Embeddings}
The reservoir dynamics for each encoded sample $\B{x}^{(k)}$ are generated by evolving the system under the sample-dependent Hamiltonian $\B{H}^{(k)}(t)$ from $t_0$ to $t_{\text{end}}$.  The evolution is governed by the unitary
\begin{align}
\B{U}^{(k)}(t) = \mathcal{T}\exp\!\Bigl[-i\!\int_{t_0}^{t}\B{H}^{(k)}(\tau)\,d\tau\Bigr],
\end{align}
where $\mathcal{T}$ denotes time-ordering, and the time-evolved state is $\ket{\psi^{(k)}(t)} = \B{U}^{(k)}(t)\ket{\psi_0}$, with $\ket{\psi_0}$ the fixed initial reservoir state. Selecting the initial state $\ket{\psi_0}$—whether a ground state, a superposition state, or with tailored pre-evolution optimization adapted to the given task—serves as a hyperparameter that can markedly influence the reservoir’s memory capacity and nonlinearity~\cite{CDLJ:24:PRR}. To extract classical features we measure, at discrete times $\{t_m\}_{m=1}^{M}$ spanning $[t_0, t_{\text{end}}]$, the set of local Pauli-$z$ operators and their pairwise correlations:
\begin{align}
\B{\hat{O}} \in 
\Bigl\{\,\B{\sigma}_i^z \,\Big|\, 1\!\le i\!\le N\Bigr\}\;
\cup\;
\Bigl\{\,\B{\sigma}_i^z\B{\sigma}_j^z \,\Big|\, 1\!\le i\!<\!j\!\le N\Bigr\},
\end{align}
where $\B{\sigma}_i^z = \ket{g_i}\!\bra{g_i} - \ket{r_i}\!\bra{r_i}$ acts on atom $i$.  These operators relate directly to the Rydberg occupations via $\B{\hat{n}}_i = \tfrac12\bigl(\B{I}-\B{\sigma}_i^z\bigr)$ and $\B{\hat{n}}_i\B{\hat{n}}_j = \tfrac14\bigl(\B{I}-\B{\sigma}_i^z-\B{\sigma}_j^z+\B{\sigma}_i^z\B{\sigma}_j^z\bigr)$. The measurement instants are uniformly spaced with step size $\Delta t$:
\begin{align}
t_m = t_0 + m\,\Delta t,\quad m=1,\dots,M,\qquad
\Delta t = \frac{t_{\text{end}}-t_0}{M}.
\end{align}
At each $t_m$ we evaluate all
\(
R = N + \binom{N}{2}
\)
observables, yielding the expectation values
\begin{align}
\expval{\B{\sigma}_i^z}_{t_m}^{(k)} &= 
\bra{\psi^{(k)}(t_m)}\B{\sigma}_i^z\ket{\psi^{(k)}(t_m)}, \\
\expval{\B{\sigma}_i^z\B{\sigma}_j^z}_{t_m}^{(k)} &= 
\bra{\psi^{(k)}(t_m)}\B{\sigma}_i^z\B{\sigma}_j^z\ket{\psi^{(k)}(t_m)}.
\end{align}
Concatenating all measurements forms the embedding vector
\begin{align}
\B{\Phi}\bigl(\B{x}^{(k)}\bigr)=\Big(
&\expval{\B{\sigma}_1^z}_{t_1}^{(k)},\;\dots,\;\expval{\B{\sigma}_N^z}_{t_1}^{(k)},\;
  \expval{\B{\sigma}_1^z\B{\sigma}_2^z}_{t_1}^{(k)},\;\dots,\;
  \expval{\B{\sigma}_{N-1}^z\B{\sigma}_N^z}_{t_1}^{(k)}, \nonumber\\
&\expval{\B{\sigma}_1^z}_{t_2}^{(k)},\;\dots,\;
  \expval{\B{\sigma}_{N-1}^z\B{\sigma}_N^z}_{t_2}^{(k)}, \;\dots,\;
  \expval{\B{\sigma}_1^z}_{t_M}^{(k)},\;\dots,\;
  \expval{\B{\sigma}_{N-1}^z\B{\sigma}_N^z}_{t_M}^{(k)}
\Big)^\top,
\end{align}
whose dimension is $D = M\bigl(N+\binom{N}{2}\bigr)$.  This spatiotemporal feature vector serves as input to the classical readout layer.

\section*{Data availability}
The datasets generated and/or analysed during the current study are available from the corresponding author upon reasonable request.

\section*{Code availability}
The source code used to support the findings of this study is available from the corresponding author upon reasonable request.


\backmatter

\clearpage

\bibliography{sn-bibliography}   

\section*{Acknowledgements}
We acknowledge the use of CUDA Quantum for this work. The views expressed are those of the authors and do not reflect the official policy or position of NVIDIA or the CUDA Quantum team..


\section*{Funding}
The work of Shehbaz~Tariq and Symeon~Chatzinotas was supported by the project Lux4QCI (GA 101091508) funded by the Digital Europe Program, and the project LUQCIA Funded by the European Union – Next Generation EU, with the collaboration of the Department of Media, Connectivity and Digital Policy of the Luxembourgish Government in the framework of the RRF program. The work of Muhammad~Talha and Hyundong~Shin is supported by the National Research Foundation of Korea (NRF) grant funded by the Korean government (MSIT) under RS-2025-00556064 and by the MSIT (Ministry of Science and ICT), Korea, under the ITRC (Information Technology Research Center) support program (IITP-2025-2021-0-02046) supervised by the IITP (Institute for Information \& Communications Technology Planning \& Evaluation)


\section*{Authors' contribution}
ST and MT contributed the idea. ST and MT developed the theory and wrote the manuscript. SC and
HS improved the manuscript and supervised the research. All authors contributed to the analysis and discussion of the results and improved the manuscript. All authors read and approved the final manuscript.

\section*{Competing interests}
The authors declare no competing interests.


\end{document}